\documentclass[]{spie}  

\usepackage{amsmath,amsfonts,amssymb}
\usepackage{graphicx}
\usepackage[colorlinks=true, allcolors=blue]{hyperref}
\usepackage{float}

\title{Gemini North Adaptive Optics (GNAO) facility overview and status updates}

\author[a]{Gaetano Sivo}
\author[b]{Julia Scharw\"achter}
\author[a]{Manuel Lazo}
\author[b]{C\'elia Blain}
\author[c,d]{Stephen Goodsell}
\author[e]{Marcos van Dam}
\author[a]{Martin Tschimmel}
\author[a]{Henry Roe}
\author[c]{Jennifer Lotz}
\author[b]{Kim Tomassino-Reed}
\author[c]{William Rambold}
\author[b]{Courtney Raich}
\author[a]{Ricardo Cardenes}
\author[b]{Angelic Ebbers}
\author[a]{Tim Gaggstatter}
\author[a]{Pedro Gigoux}
\author[b]{Thomas Schneider}
\author[b]{Charles Cavedoni}
\author[b]{Stacy Kang}
\author[b]{Stanislas Karewicz}
\author[b]{Heather Carr}
\author[b]{Jesse Ball}
\author[b]{Paul Hirst}
\author[a]{Emmanuel Chirre}
\author[b]{John White}
\author[a]{Lindsay Magill}
\author[b]{Molly Grogan}
\author[c]{Anne Jordan}
\author[f]{Suresh Sivanandam}
\author[f]{Masen Lamb}
\author[g]{Adam Muzzin}
\author[h]{Eduardo Marin}
\author[i]{Scott Chapman}
\author[j]{Jennifer Dunn}
\author[j]{Dan Kerley}
\author[j]{Jean-Pierre V\'eran}
\author[k]{Morten Andersen}
\author[l]{Franck Marchis}
\author[a]{Ruben Diaz}
\author[m]{John Blakeslee}
\author[n]{Michael Pierce}
\author[a]{Rodrigo Carrasco}
\author[a]{Hwyhyun Kim}
\author[o,p]{Anja Feldmeier-Krause}
\author[j]{Alan McConnachie}
\author[q]{James Jee}
\author[j]{Wesley Fraser}
\author[r]{Mark Ammons}
\author[t]{Christopher Packham}
\author[u]{John Bally}
\author[v]{Trent Dupuy}
\author[w]{Daniel Huber}
\author[b]{Marie Lemoine-Busserole}
\author[x]{Thomas Puzia}
\author[y]{Paolo Turri}
\author[z]{Chadwick Trujillo}
\author[c]{Janice Lee}

\affil[a]{NSF's NOIRLab - Gemini Observatory, Colina El Pino s/n Casila 603, La Serena, Chile}
\affil[b]{NSF's NOIRLab - Gemini Observatory, 670 N'Aohoku Place, Hilo, Hawaii, USA}
\affil[c]{NSF's NOIRLab - Gemini Observatory, 950 N Cherry Avenue, Tucson, Arizona, USA}
\affil[d]{Durham University, Stockton Road, Durham, UK}
\affil[e]{Flat Wavefront, 21 Lascelles Street, Christchurch, New Zealand}
\affil[f]{University of Toronto, 27 King's college Sir, Toronto, Canada}
\affil[g]{York University, 4700 Keel Street, Toronto, Canada}
\affil[h]{Keck Observatory, 65 Mamalahoa Hwy, Waimea, Hawaii, USA}
\affil[i]{Dalhousie University, 6299 South Street, Halifax, Canada}
\affil[j]{NRC Herzberg Astronomy and Astrophysics Research Centre , 5071 West Saanish Road, Victoria, Canada}
\affil[k]{European Southern Observatory, Karl-Schwarzschild-Straße 2, 85748 Garching bei München, Germany}
\affil[l]{SETI Institute, 339 N Bernardo Ave Suite 200, Mountain View, California, USA}
\affil[c]{NSF's NOIRLab, 950 N Cherry Avenue, Tucson, Arizona, USA}
\affil[n]{University of Wyoming, 1000 E University Avenue, Laramie, Wyoming, USA}
\affil[o]{University of Chicago, 5801 S Ellis Avenue, Chicago, Illinois, USA}
\affil[p]{Max Planck Institute for Astronomy, Königstuhl 17, Heidelberg, Germany}
\affil[p]{Yonsei University, 50 Yonsei-ro, Seodaemun-gu, Seoul, South Korea}
\affil[r]{Lawrence Livermore National Laboratory, 7000 East Avenue, Livermore, California, USA}
\affil[t]{University of Texas at San Antonio, 1 UTSA Circle, San Antonio, Texas, USA}
\affil[u]{University of Colorado, Boulder, Colorado, USA}
\affil[v]{University of Edinburgh, Old College, South Bridge, Edinburgh, UK}
\affil[w]{University of Hawaii, 200 W Kawili Street, Hilo, Hawaii, USA}
\affil[x]{Pontifica Universidad Catolica, Av. Libertador Bernardo O'Higgins 340, Santiago, Chile}
\affil[y]{University of British Columbia, Vancouver, Canada}
\affil[z]{Northern Arizona University, S San Francisco Street, Flagstaff, Arizona, USA}

\authorinfo{Further author information: (Send correspondence to Gaetano Sivo)\\Gaetano Sivo: E-mail: gaetano.sivo@noirlab.edu, Telephone: +56 51 2205 642\\ }

\begin{document} 
\maketitle

\begin{abstract}
The Gemini North Adaptive Optics (GNAO) facility is the upcoming AO facility for Gemini North providing a state-of-the-art AO system for surveys and time domain science in the era of JWST and Rubin operations. 

GNAO will be optimized to feed the Gemini infrared Multi Object Spectrograph (GIRMOS). While GIRMOS is the primary science driver for defining the capabilities of GNAO, any instrument operating with an f/32 beam can be deployed using GNAO. 

The GNAO project includes the development of a new laser guide star facility which will consist of four side-launched laser beams supporting the two primary AO modes of GNAO: a wide-field mode providing an improved image quality over natural seeing for a 2-arcminute circular field-of-view and a narrow-field mode providing near diffraction-limited performance over a $20 \times 20$ arcsecond square field-of-view. The GNAO wide field mode will enable GIRMOS’s multi-IFU configuration in which the science beam to each individual IFU will be additionally corrected using multi-object AO within GIRMOS. The GNAO narrow field mode will feed the GIRMOS tiled IFU configuration in which all IFUs are combined into a “super”-IFU in the center of the field. 

GNAO also includes the development of a new Real Time Controller, a new GNAO Facility System Controller and finally the development of a new AO Bench. 
We present in this paper an overview of the GNAO facility and provide a status update of each product. 
\end{abstract}


\section{INTRODUCTION}
\label{sec:intro}  
One of the main limitations for optical and near-infrared ground-based astronomy is the presence of Earth’s atmosphere. It works both to absorb light as a function of wavelength and also to blur the images. Adaptive Optics (AO) has been developed to mitigate the blurring of the atmosphere by compensating the image motion caused by atmospheric turbulence using deformable mirrors (DMs). Natural guide star (NGS) AO uses a bright natural reference source to measure the wavefront distortions introduced by turbulence and to compute the corresponding corrections to be applied to the DM. NGS AO works best in the near infrared and the reference source has to be sufficiently bright for the AO correction to be possible. A strong restriction for single source AO is the small field that can be corrected. Under good conditions, even a 20”-25” field of view (FOV) is challenging, and the quality of the correction will strongly vary within this field. Many science cases require larger FOVs and more homogeneous corrections across the FOV than achievable with single natural guide star (NGS) AO.

Science fields frequently lack sufficiently bright, nearby stars for wavefront sensing. This limitation can be overcome by using laser guide stars (LGSs). However, by passing the atmosphere twice, LGSs cannot provide any information about the tip and tilt of the wavefront, \textit{i.e.} the positional offsets seen for light from a star passing the atmosphere only once. This means that LGS AO systems typically still require an NGS for tip/tilt measurements but the requirements for this NGS can be relaxed for the benefit of larger sky coverage. Another limitation of that, in contrast to NGS, the LGS is located at a finite altitude. The turbulence layers probed with an LGS are, therefore, not identical with the ones probed with NGSs. This effect – which is called the “cone effect - can be mitigated by using multiple LGSs at different angular positions instead of a single LGS.

Various flavors of multi-guide star AO have been developed to support different types of scientific applications. We will describe the two modes that will be designed and implemented for GNAO and the added AO mode coming from the science instrument fed by GNAO. 

\begin{itemize}
    \item Laser Tomography AO (LTAO)\cite{Tallon-a-90, Lelouarn-t-00} is based on multiple LGSs and a single DM to provide high Strehl ratios over a narrow FOV. The data from the wavefront sensors are used to perform a tomographic reconstruction. Compared to single-LGS AO - as \textit{e.g.} provided by ALTAIR at Gemini-North\cite{Herriot-p-00} -, the use of multiple LGSs, probing the atmospheric turbulence along multiple lines-of-sight, helps to reduce the cone effect.
    \item Ground Layer AO (GLAO)\cite{Rigaut-p-01} uses multiple guide stars distributed over a relatively large FOV and a single DM to provide a moderate but uniform correction. The guide stars are used to determine an average correction which is then sensitive to the common area probed by the different guide stars, \textit{i.e.} the ground layer. This type of AO is most efficient if the dominant fraction of the turbulence is associated with the ground layer. This is well known for Maunakea\cite{Tokovinin-a-05} and thus well suited for GNAO. GLAO results in a seeing enhancement by concentrating the energy in the peak of the PSF. Implementations of GLAO include, e.g., GRAAL\cite{Paufique-p-16} and GALACSI\cite{Oberti-p-18} at ESO, which feed the wide-field (7.5’x7.5’) NIR imager HAWK-I and the optical large-format (1’x1’) integral field spectrograph MUSE\cite{Bacon-p-04}, respectively, and use the adaptive secondary mirror of UT4 at the VLT to apply the AO correction.
    \item Multi-Object AO (MOAO)\cite{Hammer-p-02} aims to provide an AO correction over small fields around multiple science targets selected from a large patrol field. This technique overcomes the limitations of the before-mentioned AO types to provide an efficient correction over the full patrol field. Multi-object spectroscopy (MOS) is a powerful technique for many survey-type science cases because of its multiplexing advantage. While MOS has commonly been in use under natural seeing conditions, first implementations with MOAO have only been available over recent years. The Raven instrument\cite{Lardiere-p-12} demonstrated MOAO at the 8-meter Subaru Telescope with some performance limitations because of the NGS requirements. The Gemini Infrared Multi-object Spectrograph GIRMOS\cite{Sivanandam-p-18}, built as the first-light instrument behind GNAO by a Canadian consortium, will take advantage of the GNAO LGS facility to provide an advanced implementation of MOAO, feeding multiple integral field spectrographs across the patrol field.
\end{itemize}

GNAO is developed as part of the National Science Foundation-funded “Gemini in the era of multi-messenger astronomy” (GEMMA) program, in parallel with the other two GEMMA projects on Time Domain Astronomy and Public Information and Outreach. The original design for GNAO focused on a general purpose MCAO system for imaging and spectroscopic modes provided by visiting and facility instruments\cite{Sivo-p-20}. 
During 2020, the scope of the GNAO project was refocused towards towards LTAO and GLAO using a single DM only\cite{Sivo-a-22}. This design was optimized to support the innovative first-light instrument GIRMOS\cite{Sivanandam-p-18} while reducing complexities compared to an MCAO system. 

The design of GNAO is jointly driven by the science cases developed by the GIRMOS team and the large fraction of science cases originally developed for the GNAO MCAO system that will remain feasible within the GNAO+GIRMOS scope. A potential use of GNAO with GMOS at visible wavelengths is considered an upscope \cite{Sivo-a-22}.

\section{GIRMOS: The Gemini InfraRed Multi Object Spectrograph}
The Gemini InfraRed Multi Object Spectrograph (GIRMOS) is designed as a multiplexed Near InfraRed (NIR) Integral Field Unit (IFU) spectrograph using MOAO to feed four IFUs deployable over the 2’-diameter patrol field provided by the GNAO GLAO mode. The IFUs are based on an image-slicer design offering spatial sampling options of 0.025”, 0.05”, and 0.1” per spaxel and FoVs of 1”x1”, 2”x2”, and 4”x4”, respectively. Each individual IFU will be paired with a DM to perform the MOAO correction based on the GNAO ground-layer corrected field. GIRMOS offers wavelength coverage from 0.95 to 2.4 microns at spectral resolutions of 3000 or 8000. The IFUs can be used in a tiled mode offering a total IFU FoV size that is unprecedented among current NIR IFUs at AO spatial resolutions. The resulting tiled FOV of up to 8”x8” can take advantage of the GNAO LTAO mode to achieve high Strehl ratios.
In addition to the IFU modes, GIRMOS will include a NIR imager based on a HAWAII-4RG 4K x 4K detector, which provides an 85” x 85” FOV with a spatial sampling of 0.021”/pixel. More updated informations on GIRMOS can be found in the SPIE proceeding from this conference\cite{Sivanandam-p-18}.

\section{Synergies with other facilities}
At the time of GNAO's commissioning, there will be many new facilities either in operation or close to coming online. Among the major ones are James Webb Space Telescope (JWST), the Vera C. Rubin Observatory’s Legacy Survey of Space and Time (LSST), the Nancy Grace Roman Space Telescope, and Euclid. GNAO is envisioned to complement and work in unison with these facilities.

The time domain capabilities planned for GNAO will play a key role in the follow-up of the enormous number of variable and transient objects detected with the Rubin Observatory’s LSST together with the Roman Space Telescope, and Euclid. Many of these discoveries will require rapid follow-up, both spectroscopically and photometrically, at wavelengths not covered by the Rubin Observatory/LSST camera and at higher spatial resolution (e.g., or isolating the flux from a transient against the flux of an underlying background galaxy).
  
A GNAO system with queue scheduling flexibility, fast proposal routes (like the Gemini Fast Turnaround program\cite{Andersen-p-17}), and short acquisition times for rapid target-of-opportunity triggers will be highly competitive in this context.

JWST’s launch happened on Christmas day 2021. The first data from JWST have just started to revolutionize near-to-mid-infrared astronomy. With a JWST mission lifetime goal of more than 10 years and an expected GNAO first light in the second quarter of the US fiscal year 2028, both facilities are likely going to be in operation simultaneously for several years. 

GNAO cannot compete with JWST's sensitivity, but will be substantially more flexible than JWST for rapid follow-up and long-term monitoring. The survey capabilities enabled by GNAO and GIRMOS offer major synergy potential between the two facilities. In particular, the multi-object IFU mode will provide an important multiplexing advantage, for example for spatially-resolved spectroscopy of galaxies around cosmic moon. GNAO-enabled instrument modes may provide a larger
flexibility in 
narrow-band imaging filters, or provide medium-to-high spectral resolution complementary to JWST’s capabilities. The availability of these modes depends on the backend instrument design.

\section{GNAO's Science Cases Summary}

The GNAO science team has supported the development of GNAO, to ensure the best use of the facility by the community. The team is led by Gemini/NSF’s NOIRLab and has strong participation from the Gemini Observatory community with experts in many areas of astrophysics. A collaboration with the GIRMOS team is provided through joint team members.

The development of GNAO+GIRMOS focuses on three key science capabilities that enable a wide range of science cases from solar system science to extragalactic science and cosmology. 

\begin{itemize}
    \item High Resolution facility for multi messenger events. 
    With a rapid response mode, GNAO will be a premier facility for high-angular resolution follow-up studies of gamma-ray bursts and other transients. 
    \item Powerful capabilities for Survey Science. 
    The spectroscopic multiplexing provided by GNAO+GIRMOS will provide a powerful capability for spectroscopic surveys of galactic and extragalactic sources. 
    \item Flexible for Solar System science and multi-epoch studies.
    By offering a queue-scheduled system with a non-sidereal tracking option, GNAO will have key capabilities to support a wide range of solar system science and monitoring studies. 
\end{itemize}

We present here a couple example of science cases enabled with GNAO+GIRMOS. 

\subsection{High Resolution Facility for Multi-Messenger events}

The example of the famous 2017 gravitational wave event
GW170817 has demonstrated the power of multi-messenger
astronomy as an important driver of astrophysical discoveries.
GW170817 was the first gravitational wave event detected
with an electromagnetic counterpart through a major worldwide observing campaign\cite{Abbott-a-17}. The gravitational wave signal
together with the subsequent short gamma-ray burst (GRB)
and follow-up detections across the electromagnetic spectrum
have provided unprecedented insights into the physics of
neutron star mergers, including their association with short
GRBs as well as kilonova emission from radioactive decay
during the synthesis of heavy elements. Characterizing the
kilonova emission after a short GRB is a high-priority science
driver for the GNAO narrow field mode, where high angular
resolution will improve the detection of the transient against
the background host galaxy light. As a queue-operated facility,
GNAO is being developed to offer a rapid response mode to
support this science case and early follow-up of other targets
of opportunity, such as Solar System transients or new,
unknown transients discovered with Rubin Observatory’s
Legacy Survey of Space and Time.

\subsection{Powerful capabilities for survey sicence}
GNAO, in conjunction with GIRMOS, will provide powerful
capabilities for spectroscopic surveys at high angular
resolution. The multiplexing capabilities offered by the
GIRMOS spectroscopic modes will be complementary to
many aspects of galactic and extragalactic research conducted
with the JWST. The GIRMOS spectroscopic modes are well
suited to study targets such as star clusters, galaxy nuclei and
supermassive black holes, lensed galaxies, and in the study of
galaxy evolution from low to high redshift. Two of the
primary science drivers for GIRMOS are a survey of galaxies
between redshifts of 0.7 and 2.7 and a survey of globular
clusters, taking advantage of the multi-IFU and tiled-IFU
configurations, respectively. The galaxy survey will provide
unprecedented sample sizes of high-angular-resolution IFU
data to study the role of mergers, star formation, and feedback processes in the build-up of galaxy mass around redshift 2.
Globular clusters are candidates for hosting intermediatemass black holes, which are expected to exist in the mass
range between stellar-mass and supermassive black holes but
have remained difficult to detect. The spatially resolved
spectroscopy obtained with the GIRMOS globular cluster
survey will provide new opportunities to systematically
search for observational signatures of the much sought-after
intermediate-mass black holes. 

\subsection{GNAO+GIRMOS characteristics and Image Quality requirements}
We present here the characteristics of GNAO + GIRMOS both in imaging and spectroscopic mode. 

\begin{tabular}{|c|c|c|c|c|}
\hline
     Imaging & FoV & K-band ($2.2\mu m$)  & H-band ($1.65\mu m$)  &J-band ($1.25\mu m$)   \\
     &  & performance & performance & performance \\
     \hline
     GNAO WFM (GLAO) & & & &\\ 
     60$\%$ sky coverage & $85"\times 85"$ & 120 mas FWHM & 150 mas FWHM & 200 mas FWHM \\
     \hline
     GNAO NFM (LTAO) & & & &\\ 
     60$\%$ sky coverage & $20"\times 20"$ & SR 35$\%$ & SR 20$\%$ & SR 10$\%$ \\
     \hline
     GNAO on-axis (LTAO) & & & &\\ 
     12 mag R-band limit & on-axis & SR 60$\%$ & SR 45$\%$ & SR 25$\%$ \\
     \hline
     Detector & \multicolumn{4}{|c|}{Hawaii-4RG 4k$\times$4k}\\
     \hline
     Pixel Scale & \multicolumn{4}{|c|}{0.021"}\\
     \hline
     Wavelength ($\mu$m) & \multicolumn{4}{|c|}{0.83-2.4}\\
     \hline
\end{tabular}
\section{GNAO technical configuration}

\begin{figure}[H]
  \caption{GNAO system control integration}
  \includegraphics[width=180mm,scale=1.5]{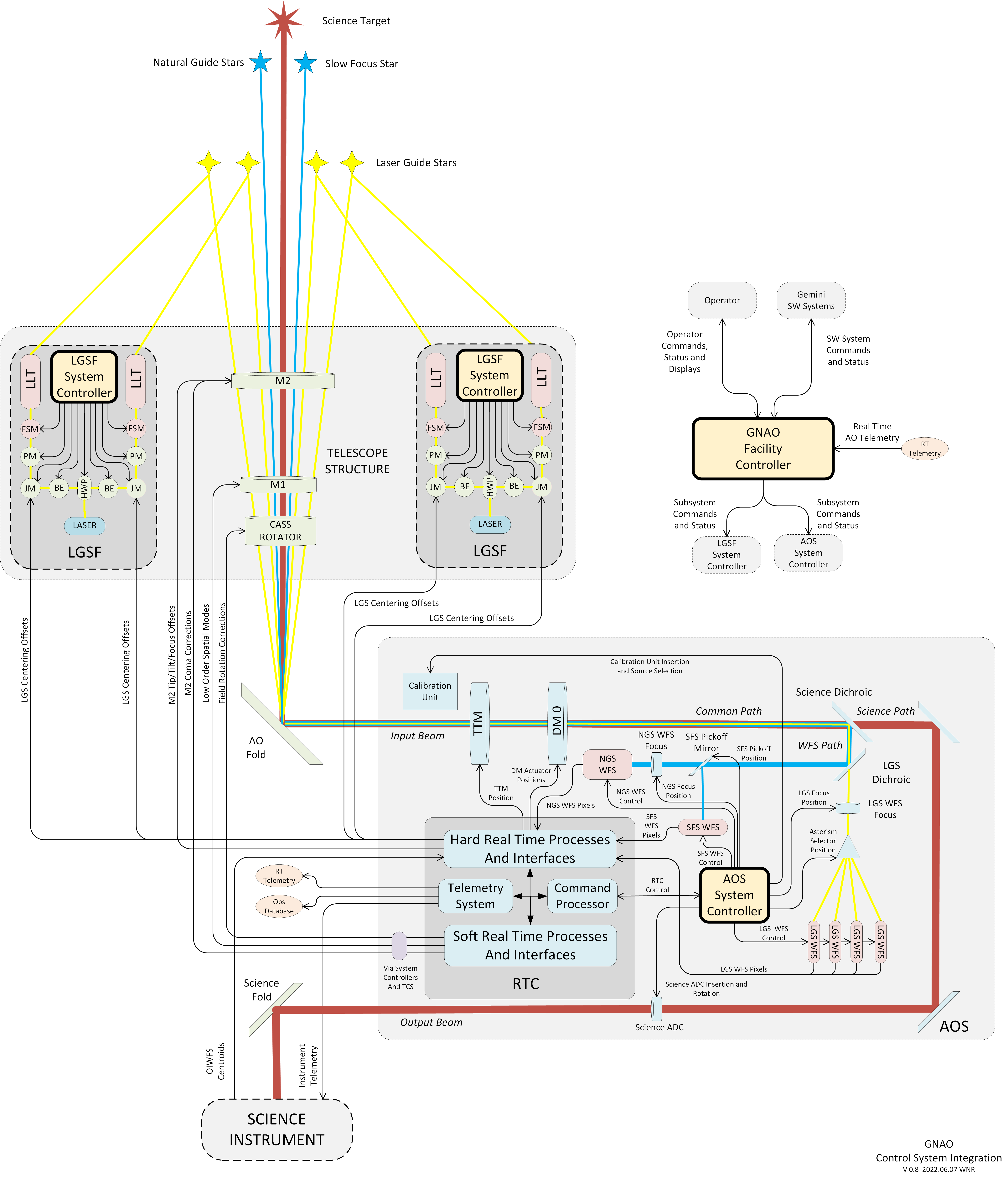}
  \label{fig:GNAOarchitecture}
\end{figure}

In this section, we present a brief overview of the different products for GNAO. The GNAO facility is divided into 4 main technical work packages: 
\begin{itemize}
    \item the laser guide star facility;
    \item the real time controller; 
    \item the system controller; 
    \item and the adaptive optics bench.
\end{itemize}
We describe these four products in the subsections below. 
In the figure \ref{fig:GNAOarchitecture}, we can see the current functional diagram of the full GNAO facility. This is not a design solution but a representation of all the functionalities required to build GNAO and perform all the different required tasks. 

\subsection{Laser Guide Star Facility}

The LGSF is responsible for generating the 4 laser guide stars that will be used for GNAO and it must also be capable of producing 4 laser guide stars in a wide field mode to support a future GLAO Adaptive Secondary Mirror system for Gemini independent of GNAO. In order to ensure the system meets requirements, the first trade study that was completed for the GNAO LGSF was the number of lasers that would be required. Due to the efficiency of the 3rd generation fiber lasers built by Toptica photonics and also due to improvements in detector technology it was determined that two Toptica lasers would be sufficient to generate four guide stars with adequate photon return for GNAO in both wide and narrow fueld mode. 

The three main components of the LGSF are the Toptica SodiumStar high-power guide star laser at 589nm, the Beam Expander and Control Node (BEaCoN), the LLTs. The 589nm lasers and LLTs will both be purchased as turn key systems through vendors, while the BEaCoN will be developed in-house. The lasers are provided by Toptica Projects GmbH.8 Officina Stellare has been selected to build the Laser Launch Telescopes. This design is similar to the LLTs that are currently in use at the European Southern Observatory (ESO) Very Large Telescope (VLT)\cite{Arsenault-p-14}. The LLTs are transmissive reverse Galilean telescopes, as opposed to the reflective off-axis parabola LLTs that are currently in use at Gemini. One key element of these LLTs is their large patrol field of 14.4 arcminutes in diameter. This allows them to support both narrow-field and wide-field constellations. The BEaCoN is the only element of the LGSF that is being developed in-house, apart from the laser coolant system.

\begin{figure}[H]
\centering
  \caption{LGSF Mechanical drawing}
  \includegraphics[width=130mm,scale=1]{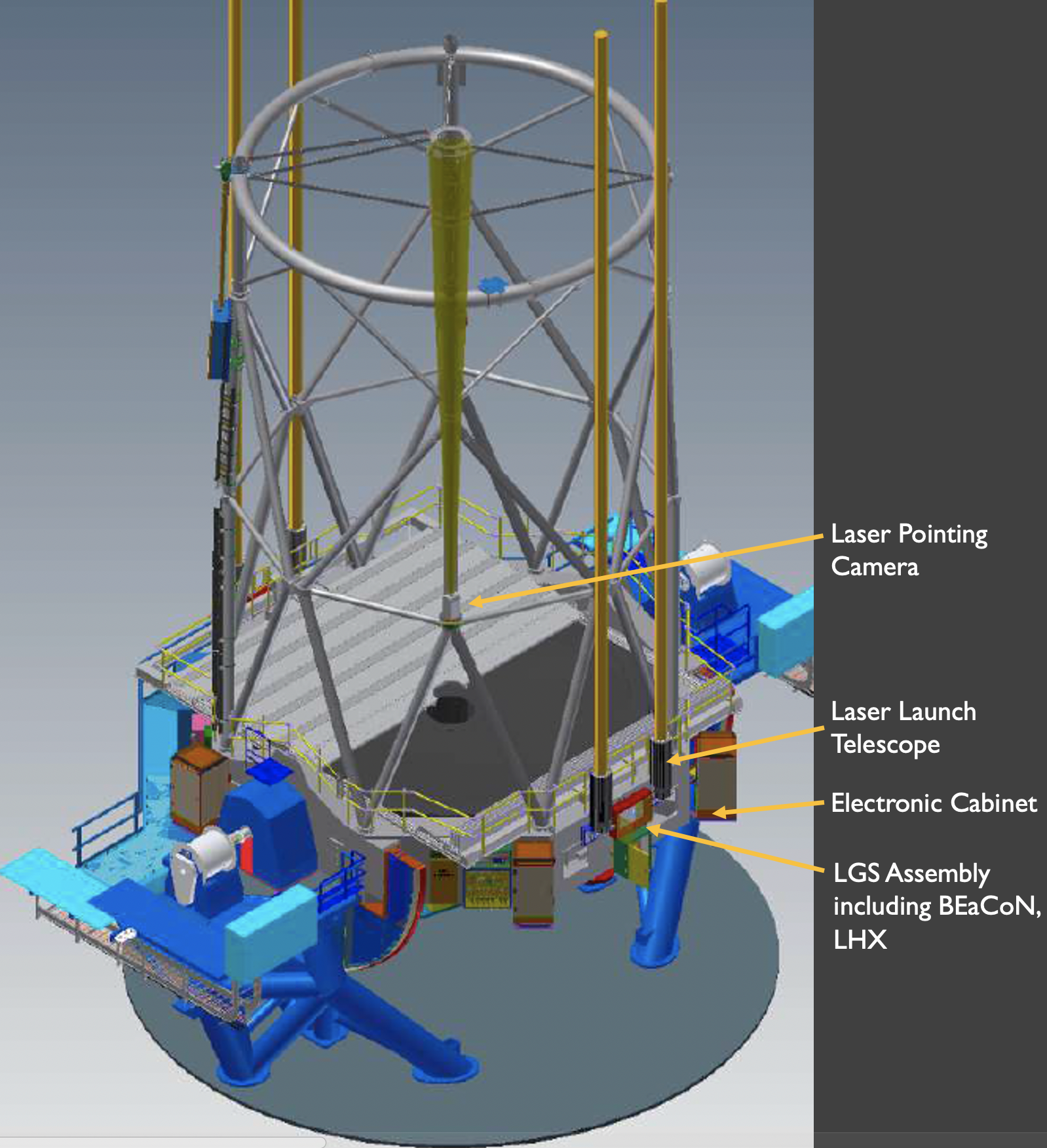}
\end{figure}

More information on the LGSF for GNAO is presented in this conference paper from Thomas Schneider\cite{Schneider-p-22}

\subsection{Real Time Controller}

The Real Time Controller for GNAO is based on the HEART toolkit\cite{Dunn-p-22}. Herzberg Extensible Adaptive Real-Time (HEART) software is a framework and collection of utility tools that can be used to create an AO Real-Time Controller (RTC). It consists of generalized RTC code blocks in modular, flexible and configurable structure that allows for development of an RTC with reduced effort, which is what everyone really wants, especially when it is an observatory that will be upgrading instruments regularly. Also, HEART comes with extensive testing of the common code base which improves the quality because it will be used across multiple projects. HEART is also unique because years ago it was decided by not tying to specific hardware it would be easier to develop if a CPU-based architecture with off-the-shelf hardware. The code is largely written in C, with some Python and JavaScript.
HEART is the baseline for several AO systems: 
\begin{itemize}
    \item GNAO on Gemini, both wide and narrow field mode. (GLAO, LTAO);
    \item NFIRAOS on Thirty Meter Telescope, MCAO;
    \item GIRMOS on Gemini, MOAO;
    \item GPI2.0 on Gemini, XAO;
    \item REVOLT on Dominion Astrophysical Observatory Telescope.
\end{itemize}
HEART is also considered for MORFEO (MCAO on ELT) and ANDES (SCAO on ELT). More details can be found in Jennifer Dunn's proceeding from this conference\cite{Dunn-p-22}. 
\subsection{System Controller}

As the current GNAO effort is ongoing, we are developing the core components for the System Control (SYSCO) facility. The System Control architecture was devised so that the different major components can reuse a basic design, divided in a number of discrete, highly configurable agents that deal with different tasks: abstracting the SYSCO-hardware interface, discrete command execution and monitoring, reactive control (subsystem coordination), and sequence execution and control. This paper describes the components providing control logic, with focus on the sequencing one, which is central to the ”One Button” philosophy guiding GNAO’s software design.
This philosophy follows from design goals of the GNAO system: improving field resolution and operating in the Gemini queue scheduled environment. From the operative point of view, GNAO strives to be a system that is simple to operate, with a single person being able to handle it without the need for an intimate knowledge of the underlying system.
To achieve these goals, the GNAO Control System is designed to support a functional decomposition of the primary actions (creating guide stars, correcting atmospheric distortion, etc), assigning them to subsystems, and then to functional components in each subsystem. Patterns of abstraction and encapsulation are used throughout the design to provide functional interfaces which ensure that each component hides the complexity and details of its underlying implementation.

In the past, Gemini's facility software has been heavily based on EPICS IOCs written in C and C++. Early on we decided against this though, first because we were looking for something more extensible. Because of that we decided to use a general-purpose language for scripting, and in order to leverage our expertise Python was the choice. More information on the detail design of the SysCo can be found in the proceeding from Ricardo Cardenes presented in this conference\cite{Cardenes-p-22}. 

\subsection{Adaptive Optics Bench}

The GNAO AOB is currently under competitive conceptual design phase studies. 3 teams have been selected to conduct the conceptual design study of the AO bench. The current plan is to held the conceptual design review early June and soon after a down selection to one team to continue the design until the delivery and integration to the telescope. More information to come later once the competitive phase has passed. 

\acknowledgments 
 
The authors would like the thank the National Science Foundation for the funding of the GEMMA
program under the Contract Support Agreement number AST-1839225. The Gemini Observatory
is operated by the Association of Universities for Research in Astronomy, Inc., under a cooperative
agreement with the NSF on behalf of the Gemini partnership: the National Science Foundation
(United States), National Research Council (Canada), CONICYT (Chile), Ministerio de Ciencia,
Tecnologìa e Innovacion Productiva (Argentina), Ministério da Ciéncia, Tecnologia e Inovaçâõ
(Brazil), and Korea Astronomy and Space Science Institute (Republic of Korea).
\bibliography{biblio_gsivo} 
\bibliographystyle{spiebib} 

\end{document}